\documentclass[review]{elsarticle}
\usepackage{graphicx}
\usepackage{amsfonts,amssymb}
\usepackage{amsmath}
\usepackage{amssymb}

\usepackage{graphics}
\usepackage{multirow}
\usepackage{amssymb}
\usepackage{caption}
\usepackage{footnote}

\usepackage[ruled]{algorithm2e}

\usepackage{textcomp,booktabs}
\usepackage[usenames,dvipsnames]{color}
\usepackage{colortbl}

\usepackage{times}
\usepackage{epsfig}
\usepackage{float}

\usepackage{rotating}

\begin{document}

\begin{frontmatter}
\title{ Cross-Modality Deep Feature Learning for Brain Tumor Segmentation}
%
%
%


\author{Dingwen Zhang\fnref{fn1}}
\ead{zhangdingwen2006yyy@gmail.com}
\author{Guohai Huang\fnref{fn1}}
\ead{huanggh666@gmail.com}
\author{Qiang Zhang\corref{cor1}\fnref{fn1}}
\ead{qzhang@xidian.edu.cn}
\author{Jungong Han\corref{cor1}\fnref{fn2}}
\ead{jungonghan77@gmail.com }
\author{Junwei Han\fnref{fn3}}
\ead{junweihan2010@gmail.com}
\author{Yizhou Yu\fnref{fn4}}
\ead{yizhouy@acm.org}

\cortext[cor1]{Corresponding author}
\fntext[fn1]{School of Mechano-Electronic Engineering, Xidian University.}
\fntext[fn2]{Computer Science Department, Aberystwyth University.}
\fntext[fn3]{School of Automation, Northwestern Polytechnical University.}
\fntext[fn4]{Deepwise AI Lab.}

\begin{abstract}
Recent advances in machine learning and prevalence of digital medical images have opened up an opportunity to address the challenging brain tumor segmentation (BTS) task by using deep convolutional neural networks. However, different from the RGB image data that are very widespread, the medical image data used in brain tumor segmentation are relatively scarce in terms of the data scale but contain the richer information in terms of the modality property. To this end, this paper proposes a novel  cross-modality deep feature learning framework to segment brain tumors from the multi-modality MRI data. The core idea is to mine rich patterns across the multi-modality data to make up for the insufficient data scale. The proposed cross-modality deep feature learning framework consists of two learning processes: the cross-modality feature transition (CMFT) process and the cross-modality feature fusion (CMFF) process, which aims at learning rich feature representations by transiting knowledge across different modality data and fusing knowledge from different modality data, respectively. Comprehensive experiments are conducted on the BraTS benchmarks, which show that the proposed cross-modality deep feature learning framework can effectively improve the brain tumor segmentation performance when compared with the baseline methods and state-of-the-art methods.

\end{abstract}

\begin{keyword}
Brain tumor segmentation, Cross-modality feature transition, Cross-modality feature fusion, Feature learning.
\end{keyword}

\end{frontmatter}

%

\section{Introduction}
\label{Introduction}

\begin{figure}
 \centering
 \includegraphics[width=8.5 cm]{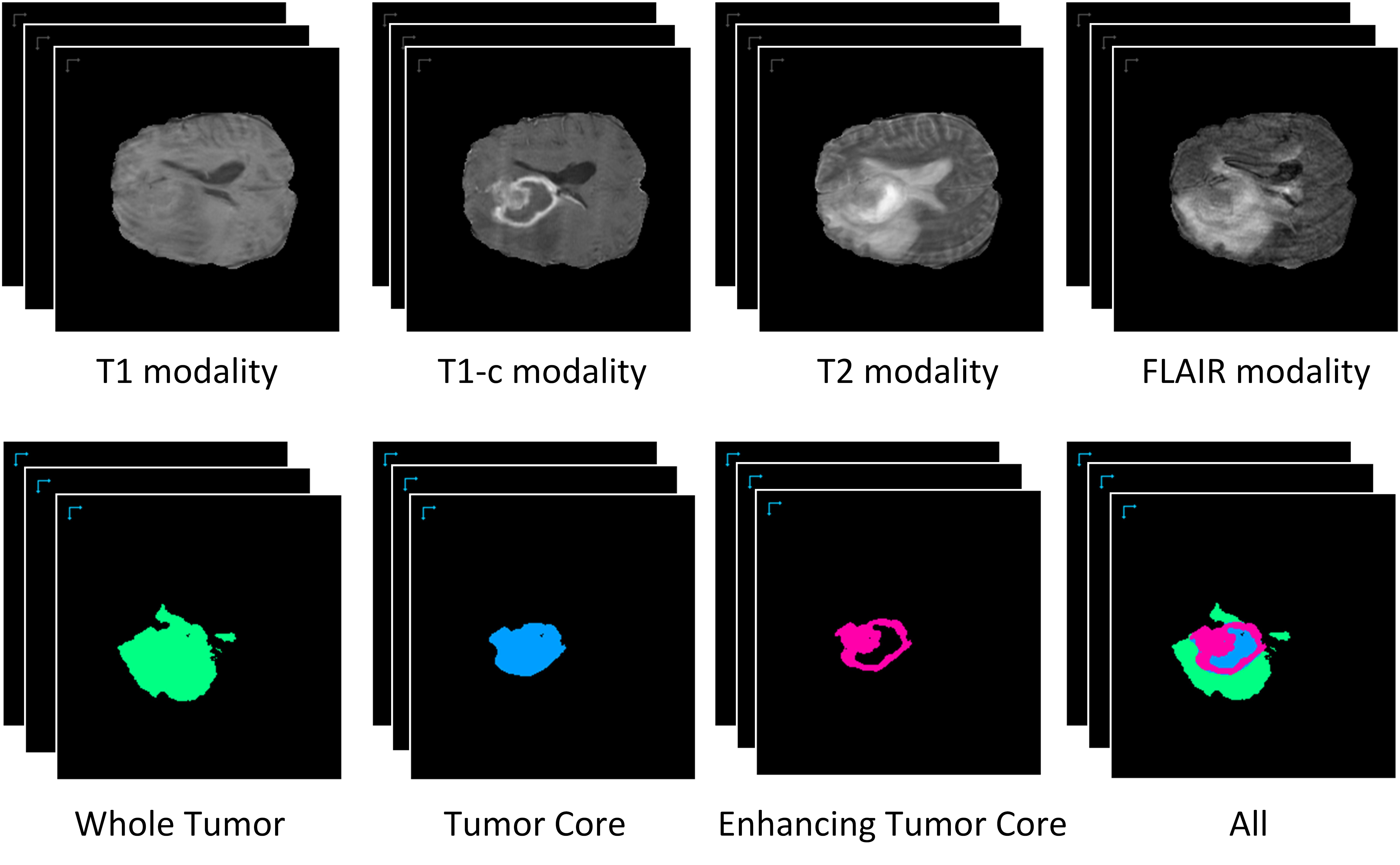}
 \caption{\textit{An illustration of the brain tumor segmentation task. The top four volume data are the multi-modality MR image data. The segmentation labels for the Whole Tumor area (WT), Tumor Core area (TC), Enhancing Tumor Core area (WT), and all types of tumor areas are shown in the bottom row. The regions without colored masks are normal areas.}} \label{fig1}
\end{figure}

As the prevailing disease with the highest mortality, the research on brain tumors has received more and more attention. In this paper, we study a deep learning-based automatic way to segment the glioma, which is called brain tumor segmentation (BTS) \cite{bakas2017segmentation}. In this task, the medical images contain four MRI modalities, which are the T1-weighted (T1) modality, contrast enhanced T1-weighted (T1c) modality, T2-weighted (T2) modality, and Fluid Attenuation Inversion Recovery (FLAIR) modality, respectively. The goal is to segment three different target areas, which are the whole tumor area, the tumor core area, and the enhancing tumor core area, respectively. An example of the multi-modality data and the corresponding tumor area labels are shown in Fig. \ref{fig1}.

With the rapid development of the deep learning technique, deep convolutional neural networks (DCNNs) have been introduced into the medical image analysis community and widely used in BTS. Given the established DCNN models, existing brain tumor segmentation methods usually consider this task as a multi-class pixel-level classification problem just as the semantic segmentation task on common RGB image data. However, by omitting the great disparity between the medical image data and the common RGB image data, such approaches would not obtain the optimal solutions. Specifically, there are two-fold distinct properties between these two kinds of data: 1) Very large-scale RGB image data can be acquired from our daily life by the smart phones or cameras. However, the medical image data are very scarce, especially for the corresponding manual annotation that requires expertise and tends to be very time consuming. 2) As a departure from the common RGB image data, the medical image data (for the investigated brain tumor segmentation task and other tasks) usually consist of multiple MRI modalities that capture different pathological properties. 

Due to the above-mentioned characteristics, BTS still has challenging issues needed to be addressed. Specifically, due to the insufficient data scale, training a DCNN model might surfer from the over-fitting issue as DCNN models usually contain numerous network parameters. This increases the difficulty of training a desired DCNN model for brain tumor segmentation. Secondly, due to the complex data structure,
directly concatenating multi-modality data to form the network input like in the previous works \cite{li2017deep,rezaei2017conditional} is neither the best choice to fully take advantage of the knowledge underlying each modality data, nor the effective strategy to fuse the knowledge from the multi-modality data.

To address these issues, this paper proposes a novel cross-modality deep feature learning framework to learn to segment brain tumors from the multi-modality MRI data. Considering the fact that the medical image data are relatively scarce in terms of the data scale but contain rich information in terms of the modality property, we propose to explore rich patterns among the multi-modality data to make up for the insufficient data scale. Specifically, the proposed cross-modality feature learning framework consists
of two learning processes: the cross-modality feature transition (CMFT) process and the cross-modality feature fusion (CMFF) process.

In the cross-modality feature transition process, we adopt the generative adversarial network learning scheme to learn useful features that can facilitate the knowledge transition across different modality data. This enables the network to mine intrinsic patterns that are helpful to the brain tumor segmentation task from each modality data. The intuition behind this process is that if the DCNN model can transit (or convert) a sample from one modality to another modality, it may capture the modality patterns of the two MRI modalities as well as the content patterns (such as the organ type and location) of this sample, while these patterns are helpful for brain tumor segmentation. In the cross-modality feature fusion process, we build a novel deep neural network architecture to take advantage of the deep features obtained from the cross-modality feature transition process and implement the deep fusion of the features captured from different modality data to predict the brain tumor areas. This is distinct from the existing brain tumor segmentation methods or the naive strategies which either 1) implement the fusion process simply at the input level, i.e., concatenating multi-modality image data as the network input, or 2) implement the fusion process at the output level, i.e., integrating the segmentation results from different modality data.

\begin{figure*}
 \centering
 \includegraphics[width=\textwidth]{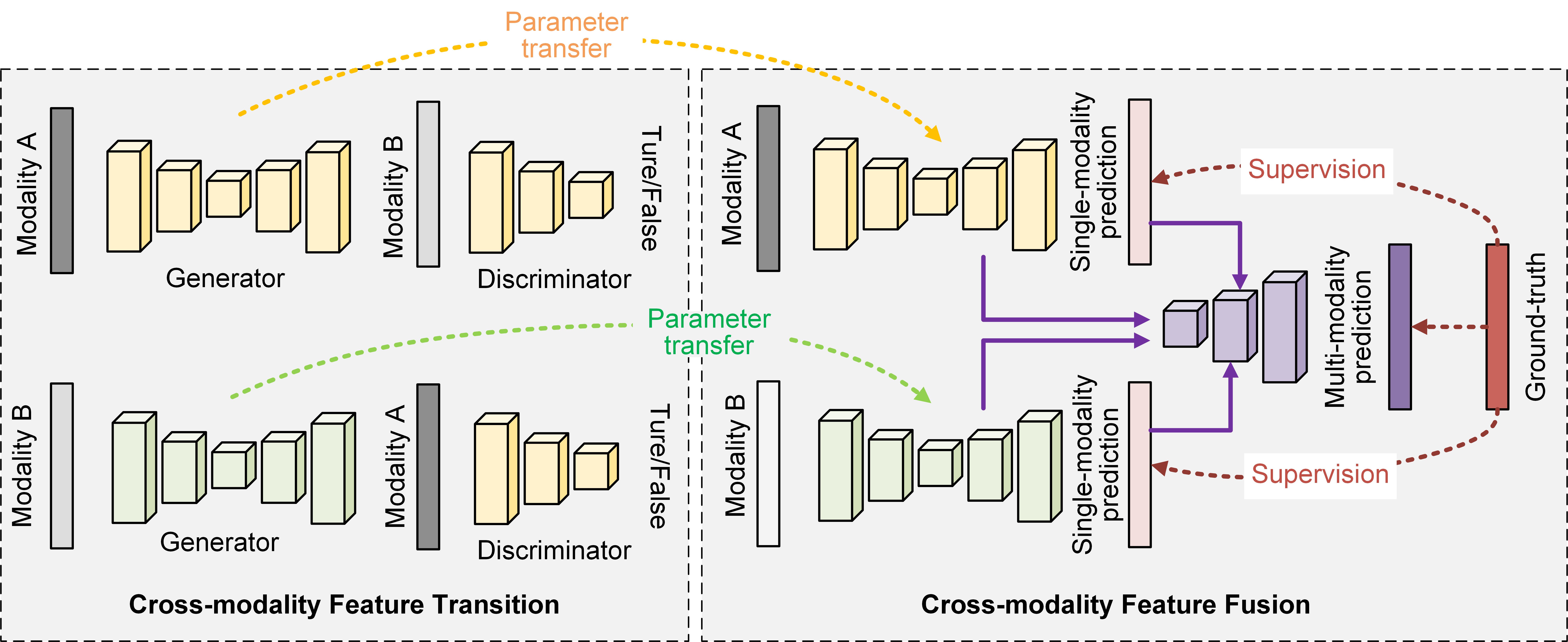}
 \caption{An illustration of the proposed cross-modality deep feature learning framework for brain tumor segmentation. To be brief and to the point, we only show the learning framework by using two-modality data. } \label{framework}
\end{figure*}

Fig. \ref{framework} illustrates the proposed learning framework briefly, from which we can observe that in the cross-modality feature transition process, we build two generators and two discriminators to transit the knowledge across the two modality data. Here the generators are used to generate one modality data from the other modality data and the discriminators aim to distinguish the generated data and the real data. While in the cross-modality feature fusion process, we adopt the generators to predict the brain tumor regions from each modality data and fuse the deep features learned from them to obtain the final segmentation results. In the fusion branch, we design a novel scheme by using the single-modality prediction results to guide the feature fusion process, which can obtain stronger feature representations during the fusion process to aid segment the desired brain tumor areas.

To sum up, this work mainly has four-fold contributions as follows:
\begin{itemize}
\item By revealing the intrinsic difference between the segmentation tasks on the medical image data and the common RGB image data, we establish a novel cross-modality deep feature learning framework for brain tumor segmentation, which consists of the cross-modality feature transition process and the cross-modality feature fusion process. 
\item We present a novel idea to learn useful feature representations from the knowledge transition across different modality data. To achieve this goal, we build a generative adversarial network-based learning scheme which can implement the cross-modality feature transition process without any human annotation.
\item For implementing the cross-modality feature fusion process, a new cross-modality feature fusion network is built for brain tumor segmentation, which transfers the features learned from the feature transition process and is empowered with the novel fusion branch to use the single-modality prediction results to guide the feature fusion process.
\item Comprehensive experiments are conducted on the BraTS benchmarks, which show that the proposed approach can effectively improve the brain tumor segmentation performance when compared with the baseline methods and the state-of-the-art methods.

\end{itemize}

\section{Related Works}
\subsection{Brain Tumor Segmentation}
Brain tumor segmentation is a hot topic in the medical image analysis and machine learning community. It has received great attention in the past few years. Early efforts in this filed designed hand-crafted features and adopted the classic machine learning models to predict the brain tumor areas. Due to the rapid development of the deep learning technique, the recent brain tumor segmentation approaches mainly apply the deep features and classifiers from the DCNN models. Based on the type of the convolutional operation used in the DCNN models, we briefly divide the existing methods into two groups, i.e., the 2D CNN-based methods and 3D CNN-based methods. The 2D CNN-based methods \cite{shaikh2017brain,islam2017fully,lopez2017dilated} apply the 2D convolutional operations and split the 3D volume samples into 2D slices or 2D patches. While the 3D CNN-based methods \cite{kamnitsas2017efficient,li2017compactness,castillo2017volumetric} apply  the 3D convolutional operations, which can take the whole 3D volume samples or the extracted sub 3D patches as the network input.

Although these deep learning-based methods can already obtain much powerful feature representation when compared to the early classical methods that are based on the hand-crafted features, they did not make full use of the multi-modality data in the feature learning process, which limits the effectiveness of the learned feature representations and the final segmentation results. Realizing this issue, Fidon et al. \cite{fidon2017scalable} proposed a multi-modal convolutional network for brain tumor segmentation, where nested network structure was designed to explicitly leverage deep features within or across modalities. Different from our approach, they did not formulate the across modality transition process and did not employ the mask guidance scheme in the feature fusion process.

\subsection{Multi-modality Feature Learning}
Multi-modality feature learning is gaining more and more attention in the recent years as the multi-modality data can provide richer information for sensing the physical world. Existing works have applied multi-modality feature learning in many computer vision-based tasks such as 3D shape recognition and retrieval \cite{bu20173d}, survival prediction \cite{yao2017deep}, RGB-D object recognition \cite{xu2017multi} and person re-identification \cite{liu2018m3l}. Among these methods, Bu et al. \cite{bu20173d} built a multi-modality fusion head to fuse the deep features learnt by a CNN network branch and a \textit{Deep Belief Network (DBN) branch}. To integrate multiple modalities and eliminate view variations, Yao et al. \cite{yao2017deep} designed a deep correlational learning module for learning informative features on the pathological data and the molecular data. In \cite{wang2015large}, Wang et al. proposed a large-margin multi-modal deep learning framework to discover the most discriminative features for each modality and harness the complementary relationship between different modalities.

Although the multi-modality feature learning technique has been applied in many computer vision tasks, it is still a under-studied issue in the research field of medical image understanding, especially for the task of brain tumor segmentation. To this end, this paper makes an early effort to build a cross-modality deep feature learning framework for brain tumor segmentation. The cross-modality feature
transition (CMFT) process and the cross-modality feature fusion (CMFF) process designed in this work are also novel to the existing multi-modality feature learning methods.

\section{The Proposed Approach}

\subsection{Cross-Modality Feature Transition}
\label{CMFT}
As shown in the left part of Fig. \ref{framework}, given modality $A$ and modality $B$, we adopt the generative adversarial learning strategy to facilitate the knowledge transition across the different modality data, which in turn captures the informative patterns from each modality data. To be specific, for each modality data, we build a generative network, i.e., the generator $G$, and a discriminative network, i.e., the discriminator $D$, to formulate the feature transition process. For achieving this goal, we apply the CycleGAN learning scheme~\cite{zhu2017unpaired} to learn the transition $G^{A}_{B} : A\rightarrow B$ and $G^{B}_{A} : B\rightarrow A$ so that
\begin{equation}
  \begin{split}
    G^{B}_{A}(G^{A}_{B}(\textbf{A}))=\textbf{A}, \\
    G^{A}_{B}(G^{B}_{A}(\textbf{B}))=\textbf{B},
  \end{split}
\end{equation}
where \textbf{A} and \textbf{B} indicates the ``real'' input sample from the modality $A$ and modality $B$, respectively. Compared with other generative adversarial learning schemes, the cycle consistency-based learning scheme adopted by the CycleGAN model has the following advantages for learning representative features: Firstly, it learns the transition $G^{A}_{B} : A\rightarrow B$ and $G^{B}_{A} : B\rightarrow A$ simultaneously, thus facilitating a better exploration of the relationship between the two modality data and maintaining the content of each modality data. Secondly, during the training process, it does not necessarily require matched modality data which might be hard to obtain in practical applications.

Besides the generators, there are also two discriminators $D_{A}$ and $D_{B}$, where  $D_{A}$ distinguish the ``fake" $A$-modality data generated by $G^{B}_{A}(\textbf{B})$ from the ``real" $A$-modality data while $D_{B}$ distinguish the ``fake" $B$-modality data generated by $G^{A}_{B}(\textbf{A})$ from the ``real" $B$-modality data. During the generative adversarial learning process, we adopt the adversarial loss to match the distribution of the generated fake data to the distribution of the ``real'' data. To this end, the adversarial loss is defined as:
\begin{equation}
  \begin{split}
    \mathcal{L}_{adv}(G^A_B,D_B) &= \mathbb{E}_B[({D_B}(\textbf{B})-1)^2] \\
    &+ \mathbb{E}_A[D_B(G^A_B(\textbf{A}))^2],
  \end{split}
\end{equation}
\begin{equation}
  \begin{split}
    \mathcal{L}_{adv}(G^B_A,D_A) &= \mathbb{E}_A[({D_A}(\textbf{A})-1)^2] \\
    &+ \mathbb{E}_B[D_A(G^B_A(\textbf{B}))^2],
  \end{split}
\end{equation}
\textit{where $\mathbb{E}_M[\tau]$ indicates the expectation of $\tau$ for all the samples from modality M.}

In addition, we also follow \cite{zhu2017unpaired} to apply the cycle consistency loss to constrain the  modality transition function $G^A_B$ and $G^B_A$ from random permution in the target modality domain. To enforce the modality transition function $G^A_B$ and $G^B_A$ to be cycle consistent, \textit{we encourage $G^A_B$ to transit the generated ``fake" $A$-modality data $G^B_A(\textbf{B})$ back to the ``real'' $B$-modality data, and similarly encourage $G^B_A$ to transit the generated ``fake" $B$-modality data $G^A_B(\textbf{A})$ back to the ``real'' $A$-modality data.} To this end, the cycle consistency loss is defined as:
\begin{equation}
  \begin{split}
    \mathcal{L}_{cyc} &= \mathbb{E}_A[||G^B_A(G^A_B(\textbf{A}))-\textbf{A}||_1] \\
                      &+ \mathbb{E}_B[||G^A_B(G^B_A(\textbf{B}))-\textbf{B}||_1].
  \end{split}
\end{equation}

By considering both the adversarial loss and the cycle consistency loss, the full learning object function of the cross-modality feature transition process becomes:
\begin{equation}
\label{ganloss}
  \arg \min_{G^A_B, G^B_A} \max_{D_A,D_B} \mathcal{L}(G^A_B,G^B_A,D_A,D_B),
\end{equation}
where
\begin{equation}
  \begin{split}
    \mathcal{L}(G^A_B, G^B_A, D_A, D_B) &= \mathcal{L}_{adv}(G^A_B, D_B) \\
                                        &+ \mathcal{L}_{adv}(G^B_A, D_A) \\
                                        &+ \lambda \mathcal{L}_{cyc}(G^A_B, G^B_A),
  \end{split}
\end{equation}
$\lambda$ is a hyper-parameter to weigh the adversarial loss and the cycle consistency loss.

\begin{figure*}[t]
 \centering
 \includegraphics[width=\textwidth]{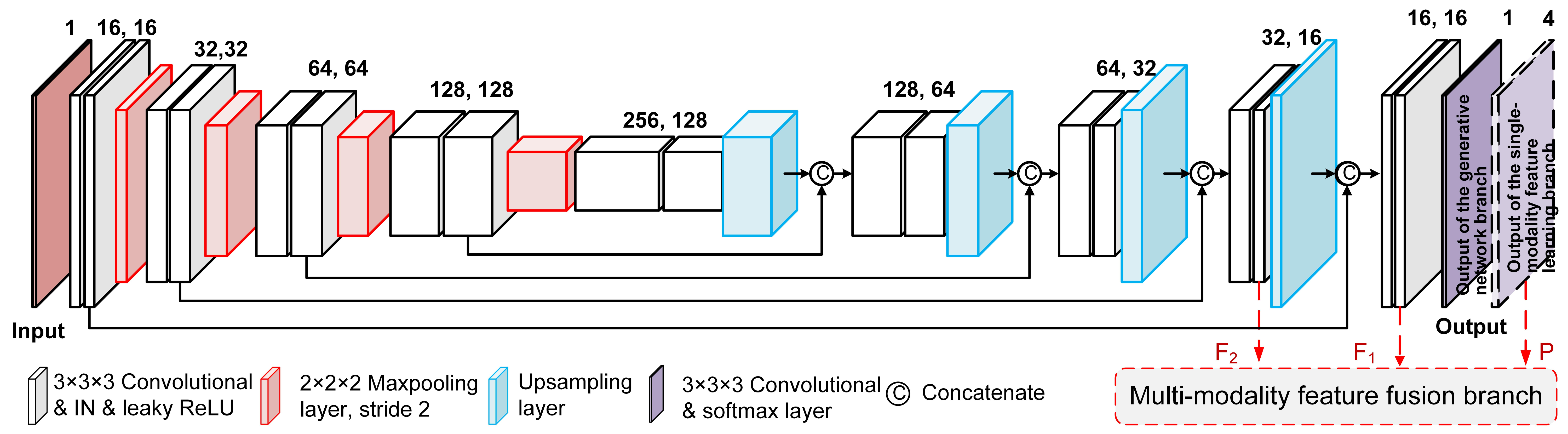}
 \caption{Illustration of the detailed architecture of the generator, where IN is short for instance normalization. Notice that this is also the architecture of the single-modality feature learning branch. The only difference between these two network branches is the last output layer, where the output of the generator is drawn in the solid line while the output of the single-modality feature learning branch is drawn in the dashed line. The deep features in the last two convolutional layers, as well as the output of the single-modality feature learning branch, are connected to the cross-modality feature fusion branch, which is annotated in red. For ease of understanding, we show the network in 2D convolution-like architecture. While we actually use the 3D convolution in network layers.} \label{unet}
\end{figure*}

~\\ \textbf{Network Architecture:} When designing the generator, we adopt a U-net architecture due to its effectiveness in both image-to-image translation \cite{wang2018perceptual} and brain tumor segmentation \cite{lopez2017dilated,shaikh2017brain,castillo2017volumetric}. Considering the training samples are in  form of 3D volumes, we adopt 3D convolutions in the network layers, thus obtaining the 3D U-net architecture. The concrete network architecture is shown in Fig. \ref{unet}. For the discriminator, we follow the existing work \cite{zhu2017unpaired} to construct it by using several convolutional layers to obtain the classification results. The concrete network architecture of the discriminator is shown in Table \ref{disc}.

\begin{table}
  \centering
  \caption{The architecture of the discriminator network branch. In the ``Input'' block, the first dimension is the number of channels and the next three dimensions are the size of the feature maps. Conv. is short for the 3D convolution, and \# filters indicates the number of filters. \textit{Notice that when learning on modality quaternions mentioned in Sec. \ref{4modality}, the number of the input channel of L1 becomes 2.}}
  \label{disc}
  \begin{tabular}{cccccc}
    \toprule
    & Type & Filter size & stride & \# filters & Input \\
    \midrule
    L1 & Conv.  & $4 \times 4 \times 4$ & 2 & 16  & $1,128,128,128$ \\
    L2 & LReLU  & -                     & - & -   & $16,64,64,64 $ \\
    L3 & Conv.  & $4 \times 4 \times 4$ & 2 & 32  & $16,64,64,64 $ \\
    L4 & INor.  & -                     & - & -   & $32,32,32,32 $ \\
    L5 & LReLU  & -                     & - & -   & $32,32,32,32 $ \\
    L6 & Conv.  & $4 \times 4 \times 4$ & 2 & 64  & $32,32,32,32 $ \\
    L7 & INor.  & -                     & - & -   & $64,16,16,16 $ \\
    L8 & LReLU  & -                     & - & -   & $64,16,16,16 $ \\
    L9 & Conv.  & $4 \times 4 \times 4$ & 2 & 128 & $64,16,16,16 $ \\
    L10 & INor. & -                     & - & -   & $128,8,8,8  $ \\
    L11 & LReLU & -                     & - & -   & $128,8,8,8  $ \\
    L12 & Conv.  & $4 \times 4 \times 4$ & 1 & 1   & $128,8,8,8  $ \\
    \bottomrule
  \end{tabular}
\end{table}

\subsection{Cross-Modality Feature Fusion}
\label{CMFF}

To implement the cross-modality feature fusion process, we establish a novel cross-modality feature fusion network for brain tumor segmentation. Equipped with the newly designed fusion branch which uses the single-modality prediction results to guide the feature fusion process, the proposed network can not only transfer the features learned from the feature transition process conveniently but also learn powerful fusion features for segmenting the desired brain tumor areas.

Given the input data from modality $A$ and $B$, the cross-modality feature fusion network contains two single-modality feature learning branches $S_A$ and $S_B$ and a cross-modality feature fusion branch $S_F$ for segmenting the desired brain tumor areas. Specifically, the single-modality feature learning branch $S_A$ takes the $A$-modality data as the input and learns representative features to predict the segmentation masks of the brain tumor areas $S_A(\textbf{A})$ as the output. Similarly, the single-modality feature learning branch $S_B$ takes the $B$-modality data as the input and learns representative features to predict the segmentation masks of the brain tumor areas $S_B(\textbf{B})$ as the output. The cross-modality fusion branch takes the deep features as well as the predicted segmentation masks of the two single-modality feature learning branches as input to learn more powerful fusion features to generate the final segmentation masks of the bairn tumor areas $S_F(\textbf{A},\textbf{B})$. To learn the cross-modality feature fusion network, we introduce the following object function:
\begin{equation}
\label{segloss}
  \arg \min_{S_A, S_B, S_F} \mathcal{L}_{seg}(S_A)+\mathcal{L}_{seg}(S_B)+\mathcal{L}_{seg}(S_F).
\end{equation}
To prevent the model from being heavily affected by the unbalance among different types of tumor areas, we follow \cite{milletari2016v} to calculate $\mathcal{L}_{seg}(S_A)$, $\mathcal{L}_{seg}(S_B)$, and $\mathcal{L}_{seg}(S_F)$ by the Dice Similarity Coefficient (DSC). Thus, for $\mathcal{L}_{seg}(S_A)$, we have
\begin{equation}
\mathcal{L}_{seg}(S_A)=1-\frac{2 \times |\textbf{Y} \cap S_A(\textbf{A})|}{|\textbf{Y}|+|S_A(\textbf{A})|},
\label{dice}
\end{equation}
where \textbf{Y} is the ground-truth annotation for the desired brain tumor areas. It goes the same for $\mathcal{L}_{seg}(S_B)$ and $\mathcal{L}_{seg}(S_F)$.

\begin{figure}
 \centering
 \includegraphics[width=12 cm]{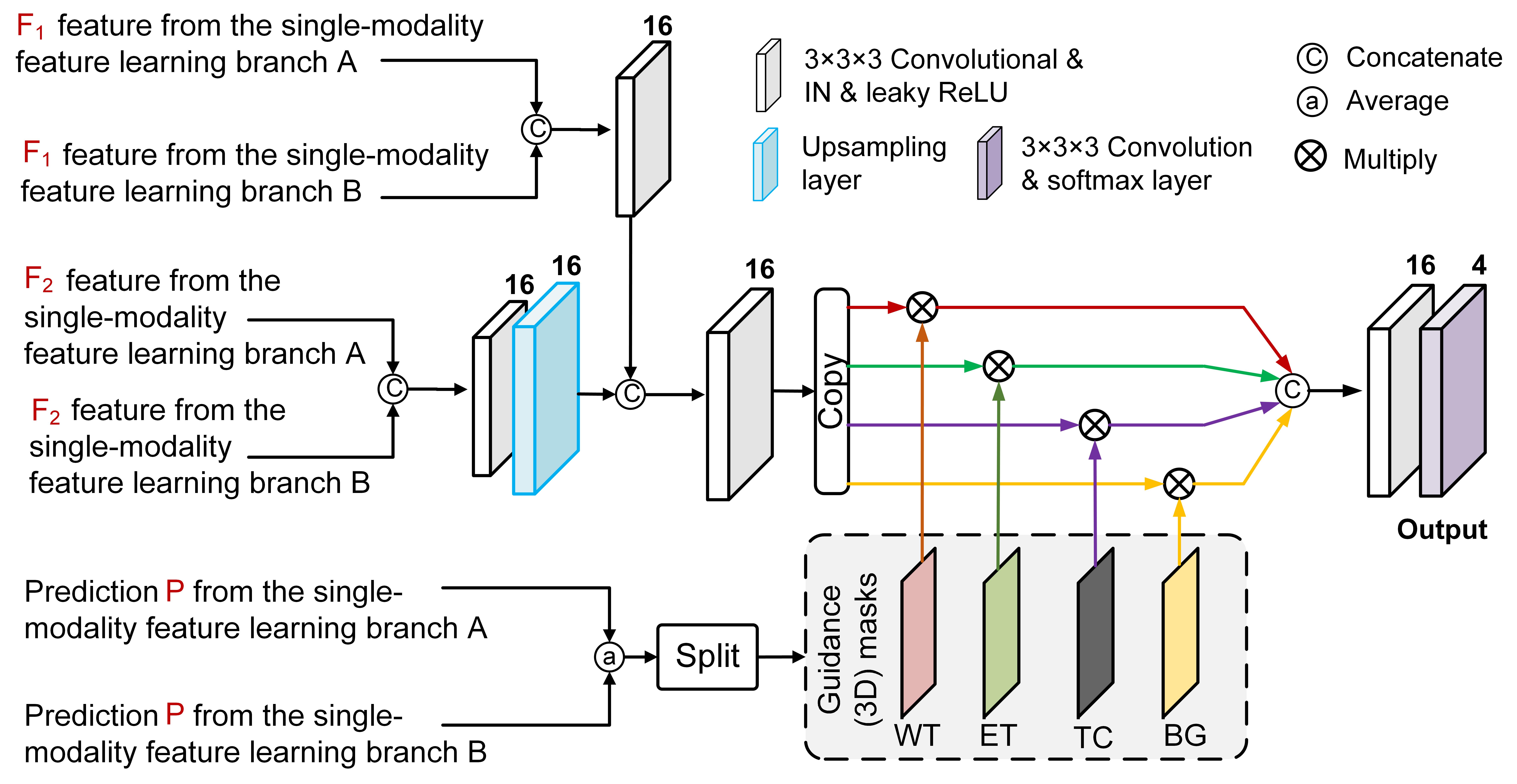}
 \caption{Illustration of the cross-modality feature fusion branch with the mask-guided feature learning scheme, where IN is short for instance normalization. For ease of understanding, we show the network in 2D convolution-like architecture. While we actually use the 3D convolution in network layers.} \label{fusion}
\end{figure}

~\\ \textbf{Network Architecture:} \textit{Although the single-modality feature learning branch does not necessarily be the same with the generator used in cross-modality feature transition, the more network layers shared by these two networks, the richer features can be conveniently transferred from the feature transition process to the feature fusion process. To this end, we adopt a quite similar network architecture to the generator $G^A_B$ (or $G^B_A$) to build the single-modality feature learning network branches $S_A$ and $S_B$ (see the right part of Fig.~\ref{framework}).} Compared to the generator, the only difference is the number of kernels set to the last convolutional layer. As shwon in Fig. \ref{unet}, the last convolutional layer of the single-modality feature learning network branch uses four convolutional kernels, while the generator only uses one convolutional kernel in the last convolutional layer. As can be seen, designing the single-modality feature learning network branch in this way could share the most network layers with the generator and thus can take full advantage of the features learned from the cross-modality feature transition process.

For fusing the knowledge from each modality data, we propose a novel cross-modality feature fusion branch. As shown in Fig. \ref{fusion}, the proposed cross-modality feature fusion branch contains several convolutional layers to fuse deep features from different layers of the two single-modality feature learning network branches. The convolutional layers are then followed by a mask-guided attention block to learn more powerful fusion features for brain tumor segmentation. Different from the conventional attention modules, such as \cite{liu2018picanet,8410621}, the attention masks in our mask-guided attention block are the segmentation masks predicted by the single-modality feature learning branches rather than those inferred from the deep feature maps from previous network layers. In other words, the attention masks in the conventional attention network blocks/modules are used to guide the network learning on its own network branch. They are learned in a bottom-up manner. In contrast, the attention masks in this work are used to guide the network learning on a different network branch and they are learned in a top-down manner.

\begin{figure}
 \centering
 \includegraphics[width=8 cm]{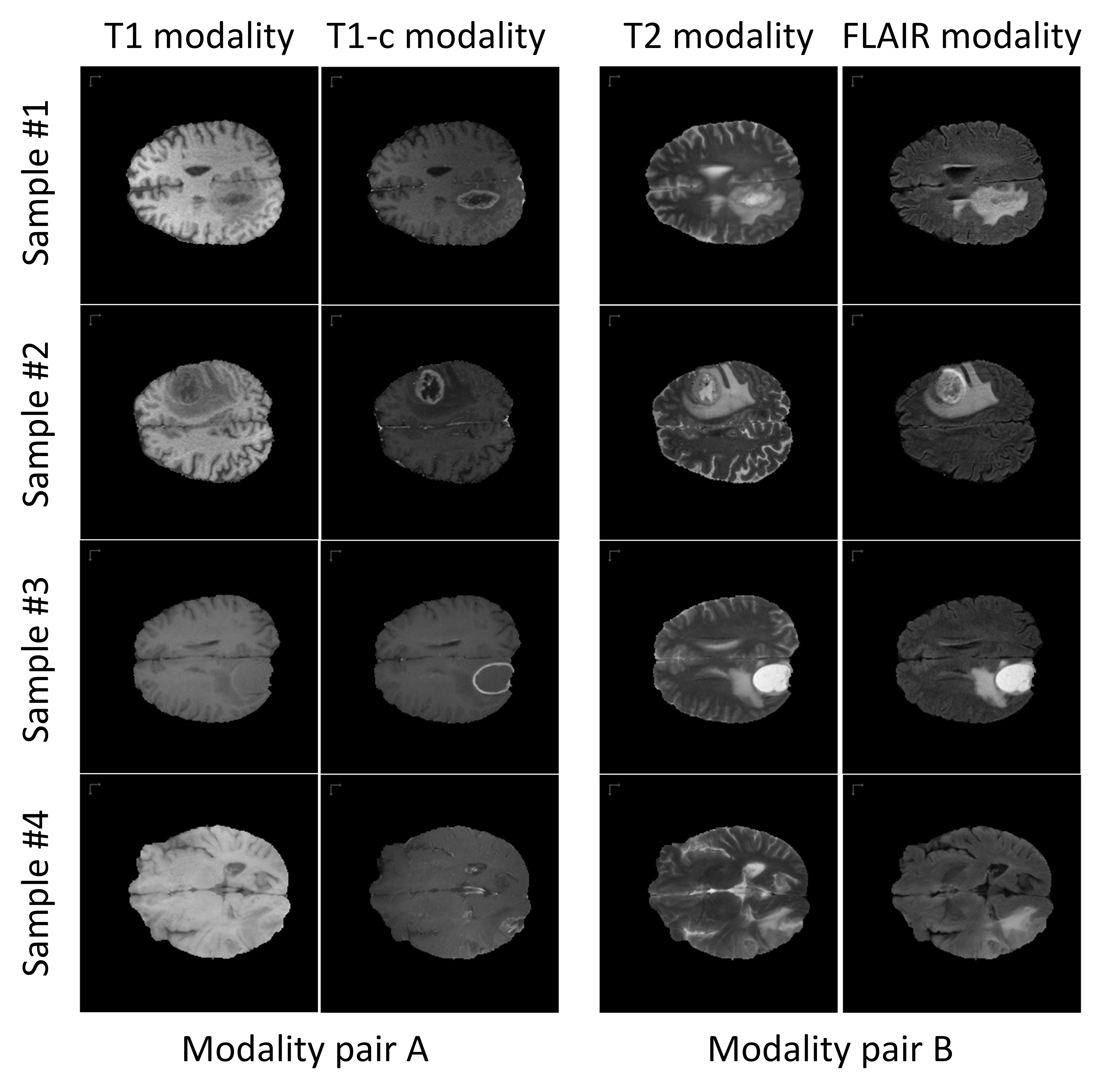}
 \caption{\textit{Examples of the modality pairs, where we use the T1 modality and T1c modality to form the modality pair $A$ while using the T2 modality and FLAIR modality to form the modality pair $B$. From the examples we can observe that the information contained within each modality pair is relatively consistent while the information contained across the different modality pairs is relatively distinct and complementary. This enables the cross-modality feature transition process to learn rich patterns.} } \label{pair}
\end{figure}

\begin{figure*}
 \centering
 \includegraphics[width=\textwidth]{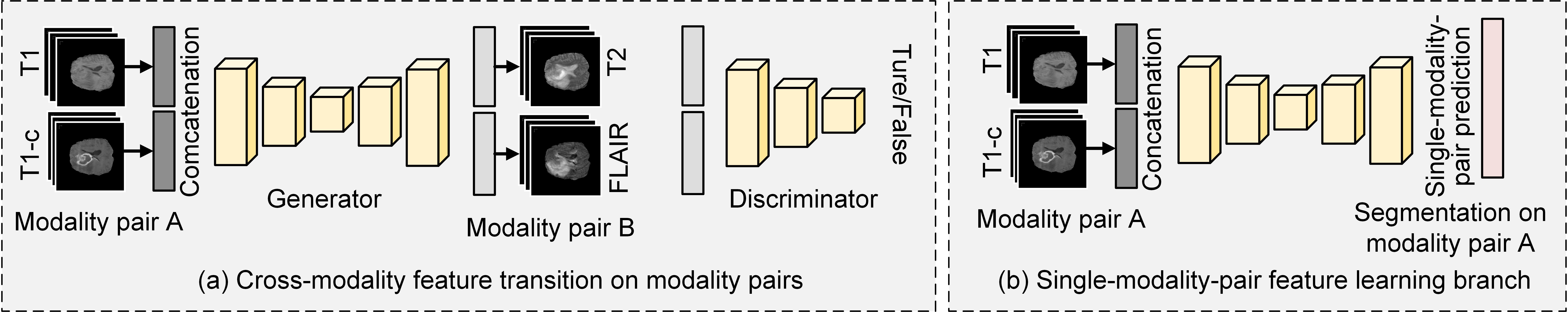}
 \caption{Illustration of the proposed strategy to extend the proposed cross-modality deep feature learning framework to work on modality quaternions. In the cross-modality feature transition process, we convert the input and output from one modality data to the concatenation of an modality pair. While in the cross-modality feature fusion process, we convert the single-modality feature learning branch to the single-modality-pair feature learning branch, which predicts the segmentation masks of each single-modality-pair. } \label{framework4}
\end{figure*}

\subsection{Learn on Modality Quaternions}
\label{4modality}
As the data used in the investigated brain tumor segmentation task usually have four modalities, i.e., the T1, T1-c, T2, and FLAIR modality (see Fig. \ref{fig1}), we also explore effective extension strategies to enable the aforementioned cross-modality deep feature learning framework to work on the modality quaternions. An naive extension is to adopt six cycGAN models, i.e, $\{G^A_B, G^B_A\}$, $\{G^A_C, G^C_A\}$, $\{G^A_D, G^D_A\}$, $\{G^B_C, G^C_B\}$, $\{G^B_D, G^D_B\}$, $\{G^C_D, G^D_C\}$, to learn the transition functions between each modality data and fuse the four single-modality feature learning branches in the cross-modality feature fusion network. Although this strategy can also learn rich feature representations from both the cross-modality feature transition and cross-modality feature fusion processes, it requires too large computational cost to implement in practice.

To this end, we propose a simple yet effective way to implement the learning framework on modality quaternions. Instead of transiting knowledge between each modality data, we implement the transition process between each modality pair. That is to say, the transition process is extended to transit knowledge from a modality pair to another modality pair. In this work, we use the T1 and T1-c modalities to form a modality pair while T2 and FLAIR modalities to form another modality pair. \textit{In this way, the information within each modality pair tends to be consistent while the information from different modality pairs tends to be distinct and complementary (see Fig. \ref{pair}), which enables the cross-modality feature transition process to learn rich patterns.} Based on this strategy, we implement the proposed approach on modality quaternions by simply converting the input data of the generators and discriminators in the CMFT process and the input data of the feature learning branch in the CMFF process to be the concatenation of two modality data, while other parts of the learning framework remain unchanged (see Fig. \ref{framework4}).

\subsection{Discussion of the Learning Framework}
As described in previous sections, the proposed learning framework contains two processes, i.e., the CMFT process and the CMFF process. In fact, these two processes can also be considered as two learning phases of a unified DCNN model. Specifically, imaging that we have a DCNN model contains two generators, two discriminators and a fusion network branch, our approach trains the two generators and the two discriminators in the first learning phase and then trains the two generators (with a modified prediction layer and loss function) together with the fusion network branch in the second learning phase. From this point of view, our proposed deep learning framework can be seen as a unified end-to-end learning model with two-phase training strategy.

Besides the two-phase training strategy, we can actually learn CMFT and CMFF simultaneously, where both Eq. \ref{ganloss} and Eq. \ref{segloss} would be introduced to form the new objective function of each training sample. However, by simultaneously learning the two generators, the two discriminators and the fusion network branch, this strategy has too much memory costs especially when exploring the 3D volume data like in this task. Thus, we choose to adopt the two-phase training strategy to implement our approach.

\section{Experiments}

\subsection{Experimental Settings}
In the BraTS 2017 and BraTS 2018 benchmark datasets, there are four modalities, i.e., T1, T1-c, T2, and FLAIR, for each patient. The BraTS 2017 benchmark has two sub-sets: a training set, which contains 285 subjects, and a validation set containing 46 subjects with hidden ground truth. The BraTS 2018 benchmark contains the same number of subjects in its training set but has 66 subjects in the validation set with hidden ground truth. When implementing the experiments on each of the benchmarks, we use the training set to train the brain tumor segmentation models while use the validation set to test the segmentation performance. We adopted the official metrics that are used by the online evaluation system of BraTS for quantitative evaluation. They are the Dice score, Sensitivity, Specificity, and the $95th$ percentile of the Hausdorff Distance (HD95).

Before training, each of the input modality data was normalized to have zero mean and unit variance. We randomly sampled patches of size $128 \times 128 \times 128$ within the brain tumor area as the inputs of both the cross-modality feature transition model and the cross-modality feature fusion model. As a trade-off between performance and memory consumption, the base number of filters in the U-Net was designed to be 16, which was increased to twice after each down-sampling layer. The Adam optimizer with an initial learning rate of $10^{-4}$ was applied to optimize the objective function, where $\lambda $ was set to be 10. When training the cross-modality feature fusion network, the pre-trained parameters of the transition mappings $G^A_B$ and $G^B_A$ were transferred to the $S_{A}$ and $S_{B}$ for further fine-tuning. The $S_{A}$ and $S_{B}$ took the same input modality data as the $G^A_B$ and $G^B_A$. The parameters of the cross-modality fusion branch are randomly initialized. We used the Adam optimizer with an initial learning rate of $10^{-4}$ and a batch size of 1 to train this network branch. All of the network branches were implemented in Pytorch on a NVIDIA GTX 1080TI GPU. \textit{It totally takes 18 hours and 57 minutes to complete the training process and the test speed is 3.2 seconds per subject.}

\begin{sidewaystable*}[p]
  \centering
  \caption{Comparison results of the proposed approach and the other baseline models on the BraTS 2017 validation set. Higher Dice and Sensitivity scores indicate the better results, while lower Hausdorff95 scores indicate the better results.}
  \label{bra17_a}
  \begin{tabular}{l|cccc|cccc|cccc}
    \hline
    \multirow{2}{*}{Method} & \multicolumn{4}{c|}{Dice} & \multicolumn{4}{c|}{Sensitivity} & \multicolumn{4}{c}{Hausdorff95} \\
    \cline{2-13}
    & ET & WT & TC & Average & ET & WT & TC & Average & ET & WT & TC & Average \\
    \hline
    \multicolumn{13}{l}{Evaluation on the key network branches with pre-train parameters from cross-modality feature transition:} \\
    $S_A$             & 0.752 & 0.799 & 0.787 & 0.779 & 0.760 & 0.787 & 0.770 & 0.772 & 3.735 & 11.640& 8.307 & 7.894 \\
    $S_B$             & 0.429 & 0.886 & 0.656 & 0.657 & 0.471 & 0.875 & 0.643 & 0.663 & 13.373& 6.072 & 10.781& 10.075\\
    $S_A+S_B$         & 0.672 & 0.864 & 0.759 & 0.765 & 0.715 & 0.834 & 0.709 & 0.753 & 7.944 & 7.032 & 7.824 & 7.600 \\
    Ours w/o MG            & 0.762 & 0.898 & 0.808 & 0.823 & 0.781 & 0.890 & 0.809 & 0.827 & 3.144 & 5.531 & 7.388 & 5.354 \\
    Ours w CA &0.765 	&0.896 	&0.799 	&0.820 	&0.776 	&0.884 	&0.753 	&0.804 	&3.402 	&4.981 	&8.066 	&5.483 \\ \hline
    \multicolumn{13}{l}{Evaluation on the cross-modality feature transition strategy based on the proposed cross-modality feature learning network:} \\
    Ours\_random & 0.725 & 0.870 & 0.754 & 0.783 & 0.766 & 0.867 & 0.774 & 0.802 & 5.826 & 6.664 & 8.748 & 7.079 \\
    Ours\_voc    & 0.725 & 0.879 & 0.778 & 0.794 & 0.768 & 0.899 & 0.766 & 0.811 & 5.202 & 6.639 & 8.642 & 6.828 \\
    Ours\_self & 0.751 	&0.898 	&0.779 &	0.809 	&0.785 	&0.884 	&0.768 &	0.812 &	3.161 &	4.775 	&7.238 &	5.058 \\
    Ours         & 0.757 & 0.900 & 0.828 & 0.828 & 0.756 & 0.904 & 0.792 & 0.817 & 3.170 & 5.155 & 6.999 & 5.108 \\

    \hline
  \end{tabular}
\end{sidewaystable*}

\subsection{Experiments on the BraTS 2017 Benchmark}

In this subsection, we evaluate the proposed approach on the BraTS 2017 benchmark. We first analyze the effect of the main network branches of the proposed learning model by conducting the experiments on the following baseline models. The first two baseline models train the single-modality-pair feature learning branches $S_A$ and $S_B$ with the input modality data \{T1,T1c\} and \{T2,FLAIR\}, respectively. The third baseline model ``$S_A+S_B$'' fuses the prediction of $S_A$ and $S_B$ by directly computing the average of the obtained segmentation maps. Then, we compare our approach with the baseline models ``Ours w/o MG'' and ``Ours w CA'' which adopt the proposed cross-modality feature fusion branch but without using the mask-guided attention block or directly using the conventional attention block \cite{Woo_2018_ECCV}. All the aforementioned baseline models are fine-tuned based on the network parameters obtained from the cross-modality feature transition process. The experimental results are reported in top rows of Table~\ref{bra17_a}.

By comparing $S_A$, $S_B$ and our approach, we can observe that simply using a single-modality-pair feature learning branch only obtains poor performance due to the inadequate modality information. The performance of $S_A+S_B$ is better than $S_B$ but worse than $S_A$, which might be caused by the large performance gap between $S_B$ and $S_A$. By comparing ``Ours w/o MG'', $S_A+S_B$, and ``Ours'' we can observe that using the proposed feature fusion branch can significantly improve the feature learning capacity of our approach and using the mask-guided attention block can further improve the segmentation accuracy. Notice that when using the conventional attention block, the network works better for the ET area but worse for the TC and WT areas, making the average performance of ``Ours w CA'' less effective than ``Ours w/o MG'' and ``Ours''.

\begin{sidewaystable*}[p]
  \centering
  \caption{Comparison results of the proposed approach and the other state-of-the-art models on the BraTS 2017 validation set. Higher Dice scores indicate the better results, while lower Hausdorff95 scores indicate the better results.}
  \label{bra17_b}
  \begin{tabular}{c|l|cccc|cccc}
    \toprule
    \multirow{2}{*}{Approach} & \multirow{2}{*}{Method} & \multicolumn{4}{c|}{Dice} & \multicolumn{4}{c}{Hausdorff95} \\
    \cline{3-10}
    & & ET & WT & TC & Average & ET & WT & TC & Average \\
    \midrule
    \multirow{6}{*}{Ensemble}
    & Kamnitsas et al. \cite{kamnitsas2017ensembles}   & 0.738 & 0.901 & 0.797 & 0.812 & 4.500 & 4.230 & 6.560 & 5.081 \\
    & Wang et al. \cite{wang2017automatic}         & 0.786 & 0.905 & 0.838 & 0.843 & 3.282 & 3.890 & 6.479 & 5.097 \\
    & Isensee et al. \cite{isensee2017brain}    & 0.732 & 0.896 & 0.797 & 0.808 & 4.550 & 6.970 & 9.480 & 7.000 \\
    & Jungo et al. \cite{jungo2017towards}       & 0.749 & 0.901 & 0.790 & 0.813 & 5.379 & 5.409 & 7.487 & 6.092 \\
    & Hu et al. \cite{hu20173d}              & 0.650 & 0.850 & 0.700 & 0.733 & 17.980& 25.240& 21.450& 21.557\\
    & Casamitjana et al. \cite{casamitjana2017cascaded} & 0.714 & 0.877 & 0.637 & 0.743 & 5.434 & 8.343 & 11.173& 8.317 \\
    \hline
    \multirow{6}{*}{Single prediction}
    & Islam et al. \cite{islam2017multi}  & 0.689 & 0.876 & 0.761 & 0.775 & 12.938& 9.820 & 12.361& 11.706\\
    & Jesson et al. \cite{jesson2017brain} & 0.713 & 0.899 & 0.751 & 0.788 & 6.980 & 4.160 & 8.650 & 6.597 \\
    & Roy et al. \cite{roy2018recalibrating}    & 0.716 & 0.892 & 0.793 & 0.800 & 6.612 & 6.735 & 9.806 & 7.718 \\
    & Pereira et al. \cite{pereira2019adaptive} & 0.733 & 0.895 & 0.798 & 0.809 & 5.074 & 5.920 & 8.947 & 6.647 \\
    & Castillo et al. \cite{castillo2017brain} & 0.710 & 0.880 & 0.680 & 0.757 & 6.120 & 9.630 & 11.380& 9.043 \\
    & Ours            & 0.762 & 0.898 & 0.823 & 0.828 & 3.170 & 5.155 & 6.999 & 5.108 \\
    \bottomrule

  \end{tabular}
\end{sidewaystable*}

In addition, we also conducted the ablation study by implementing three baseline models which directly train the CMFF network to obtain the segmentation results without the CMFT process. The first baseline ``Ours\_random'' used the random values to initialize the CMFF network, while the second baseline ``Ours\_voc'' used the parameters pre-trained on the PASCAL VOC segmentation dataset \cite{everingham2010pascal}\footnote{PASCAL VOC segmentation dataset is a large-scale image set that consists of RGB images and the corresponding segmentation annotation.} to initialize the CMFF network. To facilitate the parameter transferring between the 2D image data and 3D volume data, we first trained a 2D-Unet on the PASCAL VOC segmentation dataset and then extended its convolution kernels to 3D convolution kernels as in~\cite{Carreira2017Quo}. For the third baseline ``Ours\_self'', we replaced the proposed CMFT process by a self reconstruction-based feature learning process that learns patterns by reconstructing the input data.

The experimental results are reported in bottom rows of Table~\ref{bra17_a}. From the comparison results, we can observe that 1) due to the inadequate of medical imaging data, directly training the DCNN models with random parameter initialization is not able to achieve satisfying learning performance; 2) while using the large-scale RGB image data (together with the segmentation annotation) still cannot solve this problem because of the large domain gap; and 3) the proposed cross-modality feature transition process can learn informative features from the medical imaging data without using any human annotation, which also works better than the self reconstruction-based learning strategy.

Next, we compare the proposed approach with several state-of-the-art methods, which include six ensemble methods and five single prediction methods. \textit{The ensemble methods integrate multiple deep brain tumor segmentation models that are trained from different views or different training sub-sets to obtain the predicted segmentation masks for each test data, while the single prediction methods only apply one deep model to fulfill the brain tumor segmentation task.} Thus, the ensemble methods can usually obtain better performance but with higher complexity both in computational cost and time consumption. The quantitative results are reported in Table~\ref{bra17_b}. From Table~\ref{bra17_b}, we can observe that as a single prediction method\footnote{\textit{Although our model has a cross-modality feature transition process and a cross-modality feature fusion process, the cross-modality feature transition process only learns features and does not predict segmentation results. In other words, our segmentation results are predicted by the cross-modality feature fusion process only rather than the combination of the segmentation results obtained by both processes. Thus, our approach is considered as a single prediction method rather than an ensemble method.}}, our proposed approach outperforms all the state-of-the-art single prediction methods both in terms of Dice score and Hausdorff95. More encouragingly, our approach can also obtain better performance than most (nine out of ten) ensemble methods. Thus, the comparison results in Table~\ref{bra17_b} demonstrate the effectiveness of the proposed approach.

\begin{sidewaystable*}[p]
  \centering
  \caption{Comparison results of the proposed approach and the other baseline models on the BraTS 2018 validation set. Higher Dice and Sensitivity scores indicate the better results, while lower Hausdorff95 scores indicate the better results.}
  \label{bra18_a}
  \begin{tabular}{l|cccc|cccc|cccc}
    \toprule
    \multirow{2}{*}{Method} & \multicolumn{4}{c|}{Dice} & \multicolumn{4}{c|}{Sensitivity} & \multicolumn{4}{c}{Hausdorff95} \\
    \cline{2-13}
     & WT & ET & TC & Average & WT & ET & TC & Average & WT & ET & TC & Average \\
    \midrule
    \multicolumn{13}{l}{Evaluation on the key network branches with pre-train parameters from cross-modality feature transition:} \\
    $S_A$            & 0.786 & 0.807 & 0.812 & 0.802 & 0.862 & 0.816 & 0.826 & 0.835 & 4.350 & 10.060& 9.670 & 8.027 \\
     $S_B$            & 0.444 & 0.898 & 0.704 & 0.682 & 0.468 & 0.905 & 0.708 & 0.694 & 11.164& 5.212 & 9.895 & 8.757 \\
     $S_A+S_B$          & 0.721 & 0.873 & 0.797 & 0.797 & 0.770 & 0.863 & 0.782 & 0.805 & 4.255 & 6.301 & 7.323 & 5.960 \\
     Ours w/o MG& 0.781 & 0.900 & 0.822 & 0.834 & 0.794 & 0.916 & 0.836 & 0.849 & 3.948 & 4.449 & 7.348 & 5.248 \\
     Ours w CA &  0.788 	&0.901 	&0.833 	&0.841 	&0.836 	&0.921 	&0.829 	&0.862 	&3.788 	&5.140 	&6.265 	&5.064 \\ \hline

     \multicolumn{13}{l}{Evaluation on the cross-modality feature transition strategy based on the proposed cross-modality feature learning network:} \\
    Ours\_random & 0.755 & 0.873 & 0.771 & 0.800 & 0.772 & 0.886 & 0.811 & 0.823 & 5.340 & 6.084 & 9.082 & 6.835 \\
    Ours\_voc    & 0.760 & 0.896 & 0.785 & 0.814 & 0.744 & 0.911 & 0.747 & 0.800 & 3.300 & 4.700 & 8.427 & 5.476 \\
    Ours\_self &0.767 	&0.898 	&0.832 	&0.833 	&0.767 	&0.886 	&0.821 	&0.825 	&3.029 	&5.272 	&6.296 	&4.865 \\
    Ours        & 0.791 & 0.903 & 0.836 & 0.843 & 0.846 & 0.919 & 0.835 & 0.867 & 3.992 & 4.998 & 6.369 & 5.120 \\

    \bottomrule
  \end{tabular}
\end{sidewaystable*}


\subsection{Experiments on the BraTS 2018 Benchmark}
\begin{sidewaysfigure*}[p]
  \centering
  \includegraphics[width=20 cm]{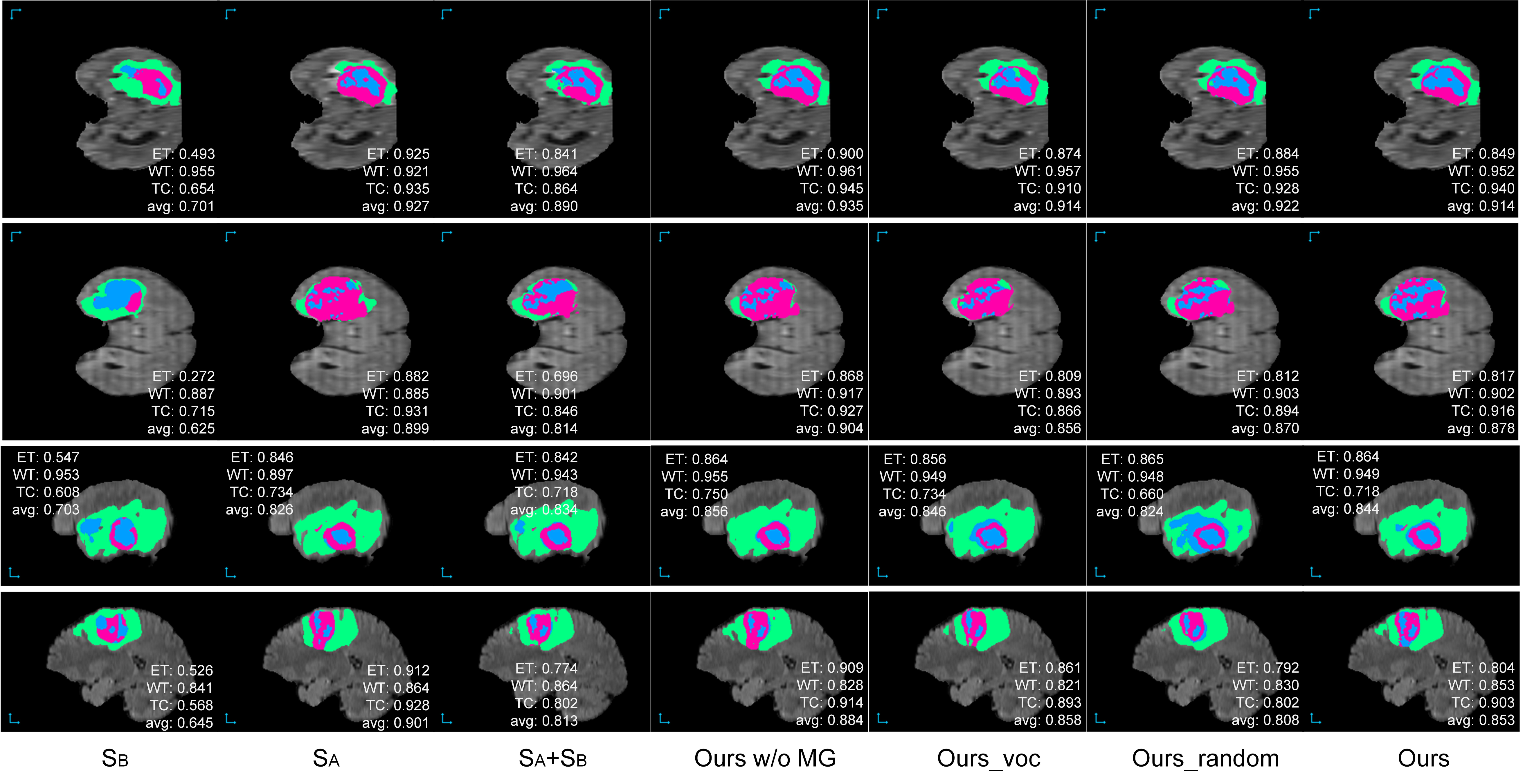}
  \caption{Examples of the segmentation results of the proposed approach as well as the compared baseline methods on the BraTS 2018 validation set. As the ground-truth segmentation annotation is not acquirable, we annotate the dice score for the segmented tumor regions on each test sample instead of showing the ground-truth segmentation annotation. The WT, TC, and ET areas are masked in green, blue, and purple, respectively. }
  \label{brats_18}
\end{sidewaysfigure*}

\begin{figure*}
 \centering
 \includegraphics[width=\textwidth]{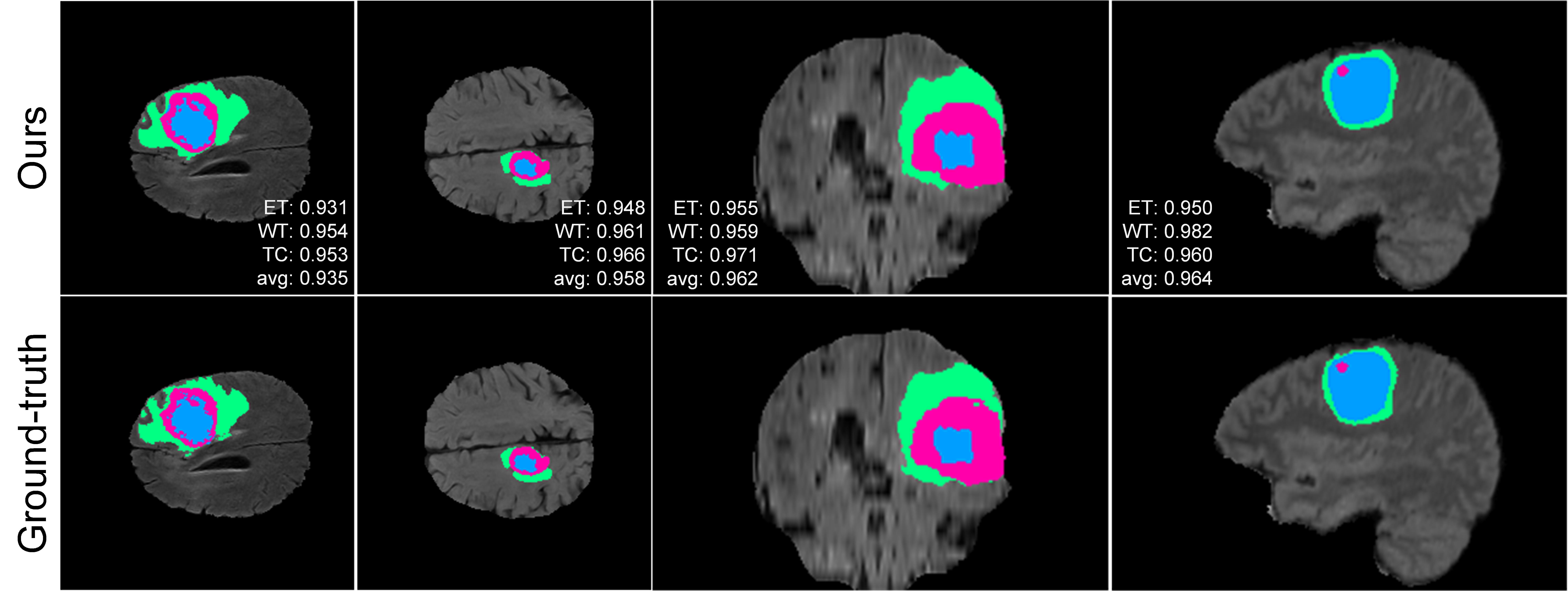}
 \caption{\textit{Comparison of the segmentation results and the ground-truth annotation on the BraTS 2018 training set. Notice that the average Dice score on the BraTS 2018 training set is 0.886, which is moderately higher than the Dice score on the BraTS 2018 validation set. The WT, TC, and ET areas are masked in green, blue, and purple, respectively. } } \label{brats_18t}
\end{figure*}

\begin{sidewaysfigure*}[p]
  \centering
  \includegraphics[width=16 cm]{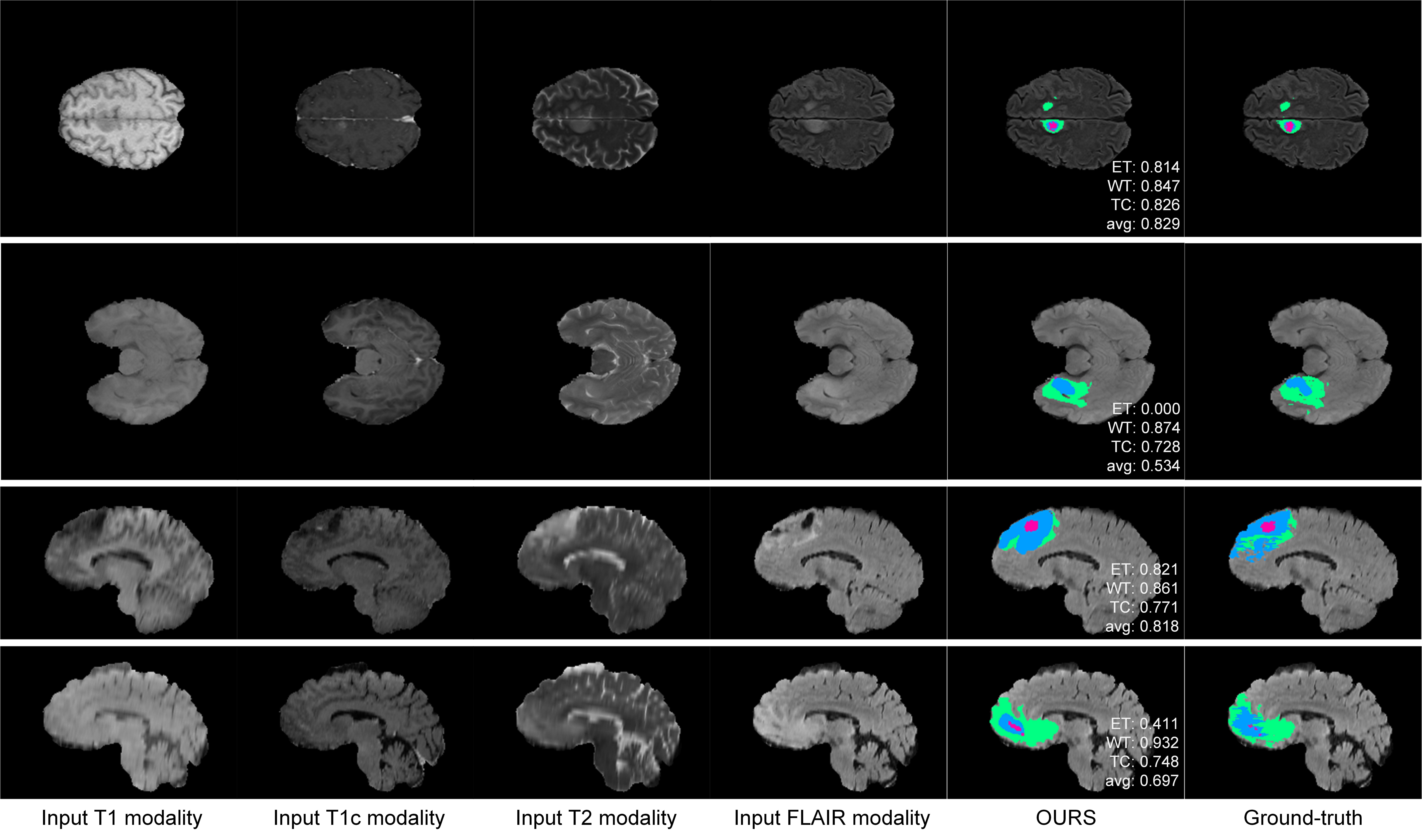}
  \caption{\textit{Examples of the failure cases on the BraTS 2018 training set, where the WT, TC, and ET areas are masked in green, blue, and purple, respectively. The first example is a failure case in the HGG subjects, which is mainly due to the inaccurate tumor boundaries. The other examples are from the LGG cases, where the absent ET area, discontinuous tumor regions and ragged tumor contours make the model hard to predict.} }
  \label{fail}
\end{sidewaysfigure*}

On the larger-scale BraTS 2018 benchmark, we first compare the proposed approach with five baseline models, including ``$S_A$'', ``$S_B$'', ``$S_A+S_B$'', ``Ours w/o MG'', and ``Ours w CA'' to analyze the effect of the main network branches designed in our learning framework. The experimental results are reported in top rows of Table \ref{bra18_a}. Being consistent with the results on the BraTS 2017 benchmark, there is obvious performance gap between ``$S_A$'' and  ``$S_B$'' and the straightforward fusion strategy ``$S_A+S_B$'' can only obtain performance better than ``$S_B$'' but worse than ``$S_A$''. Compared to ``$S_A+S_B$'', our approach obtains 4.6\% performance gain (in terms of the Dice score), which demonstrates that the feature fusion branch proposed by our approach plays an important role in fusing informative features and predicting accurate tumor areas. Notice that ``Ours w CA'' obtains better performance than ``Ours w/o MG'' on this dataset. But its performance is still worse than ``Ours''. Some examples of the comparison results on the BraTS 2018 validation set are shown in Fig.\ref{brats_18}. \textit{For better understanding the segmentation results, we also shown examples of our approach on the BraTS 2018 training set with the corresponding ground-truth annotations (see Fig.\ref{brats_18t}). Besides, we also study the failure cases in Fig. \ref{fail}, from which we can observe that the main challenges to our approach are the LGG cases when the ground-truth tumor areas are with absent ET area, discontinuous tumor regions, or ragged tumor contours. }

To evaluate the effectiveness of the proposed CMFT process, we also compare our approach with the ``Ours\_random'', ``Ours\_voc'', and ``Ours\_self'' baselines. The experimental results are reported in bottom rows of Table \ref{bra18_a}, from which we can observe obvious performance gain when compare our approach to the aforementioned baseline methods. Some examples of the comparison results are shown in Fig. \ref{brats_18}, which can better illustrate the advantage of our approach. \textit{In addition, to verify the effectiveness of our strategy to build the modality pairs as described in Sec. \ref{4modality}, we further implement a baseline model which constructs the modality pair A by using the T1 modality and FLAIR modality and modality pair B by using the T2 modality and T1-c modality. Based on our experiment, this baseline obtains 0.822 Dice score, 0.844 sensitivity, and 5.789 Hausdorff Distance on the BraTS 2018 dataset. The comparison between this baseline and the proposed approach demonstrates the effectiveness of our approach in making the information contained within each modality pair relatively consistent and the information contained across the different modality pairs relatively distinct and complementary.}

Finally, we compare the proposed approach with other state-of-the-art methods on the BraTS 2018 benchmark, which include three ensemble models \cite{myronenko20183d,Isensee2018No,puch2018global} and three single prediction models \cite{chandra2018context,ma2018automatic,chen2018s3d}. It is worth mentioning that as different works adopt various ways to obtain their ensemble models and the concrete processes for obtaining the ensemble model are not clear to us, it is hard to implement an ensemble model that could compare with the existing ensemble models fairly. However, from the experimental results reported in Table \ref{bra18_b}, we can observe that our single-prediction model has already achieved better performance when compared to the ensemble models of \cite{puch2018global,hua2018multimodal}. When compared to the state-of-the-art single prediction models, our approach also obtains the outperforming performance both in terms of Dice score and Hausdorff95. Thus, we believe the above experiments have already demonstrated the effectiveness of the proposed approach.

\begin{sidewaystable*}[p]
  \centering
  \caption{Comparison results of the proposed approach and the other state-of-the-art models on the BraTS 2018 validation set. Higher Dice scores indicate the better results, while lower Hausdorff95 scores indicate the better results.}
  \label{bra18_b}
  \begin{tabular}{c|l|cccc|cccc}
    \toprule
    \multirow{2}{*}{Approach} & \multirow{2}{*}{Method} & \multicolumn{4}{c|}{Dice} & \multicolumn{4}{c}{Hausdorff95} \\
    \cline{3-10}
    & & ET & WT & TC & Average & ET & WT & TC & Average\\
    \midrule
    \multirow{3}{*}{Ensemble}
    & Myronenko A. \cite{myronenko20183d}    & 0.823 & 0.910 & 0.867 & 0.866 & 3.926 & 4.516 & 6.855 & 5.099 \\
    & Isensee et al. \cite{Isensee2018No}  & 0.809 & 0.913 & 0.863 & 0.861 & 2.410 & 4.270 & 6.520 & 4.400 \\
    & Puch et al. \cite{puch2018global}     & 0.758 & 0.895 & 0.774 & 0.809 & 4.502 & 10.656& 7.103 & 7.420 \\
    \hline
    \multirow{3}{*}{Single prediction}
    & Chandra et al. \cite{chandra2018context}    & 0.767 & 0.901 & 0.813 & 0.827 & 7.569 & 6.680 & 7.630 & 7.293 \\
    & Ma et al. \cite{ma2018automatic}   & 0.743 & 0.872 & 0.773 & 0.796 & 4.690 & 6.120 & 10.400& 7.070 \\
    & Chen et al. \cite{chen2018s3d}        & 0.733 & 0.888 & 0.808 & 0.810 & 4.643 & 5.505 & 8.140 & 6.096 \\
    & Ours                & 0.791 & 0.903 & 0.836 & 0.843 & 3.992 & 4.998 & 6.369 & 5.120 \\
    \bottomrule

  \end{tabular}
\end{sidewaystable*}

\section{Conclusion}

In this work, we have proposed a novel cross-modality deep feature learning framework for segmenting brain tumor areas from the multi-modality MR scans. Considering that the medical image data for brain
tumor segmentation are relatively scarce in terms of the data scale but containing the richer information in terms of the modality property, we propose to mine rich patterns across the multi-modality data to make up for the insufficiency in data scale. The proposed learning framework consists of a cross-modality feature
transition (CMFT) process and a cross-modality feature fusion (CMFF) process. \textit{By building a generative adversarial network-based learning scheme to implement the cross-modality feature transition process, our approach is able to to learn useful feature representations from the knowledge transition across different modality data without any human annotation. While the cross-modality feature fusion process transfers the features learned from the feature transition process and is empowered with the novel fusion branch to guide a strong feature fusion process.}
 Comprehensive experiments are conducted on two BraTS benchmarks, which demonstrate the effectiveness of our approach when compared to baseline models and state-of-the-art methods. \textit{To our knowledge, one limitation of this work is the current learning framework requires that the network architectures of the modal generator and the segmentation predictor be almost the same. To address this inconvenience, one potential future direction is to introduce the knowledge distillation mechanism \cite{cho2019efficacy,hao2019spatiotemporal,xu2019lightweightnet} to replace the simple parameter transfer process. }


%



\section*{Acknowledgment}

This work was supported in part by the National Science Foundation of China under Grants 61876140 and 61773301, the Fundamental Research Funds for the Central Universities under Grant JBZ170401, and the China Postdoctoral Support Scheme for Innovative Talents under Grant BX20180236.





\bibliographystyle{elsarticle-num}
\bibliography{BTSegbib}

%

%




\end{document}